\documentclass[sigconf]{acmart}

\AtBeginDocument{%
  }

\setcopyright{acmcopyright}

\copyrightyear{2022}
\acmYear{2022}
\setcopyright{acmcopyright}\acmConference[CIKM '22]{Proceedings of the 31st ACM
International Conference on Information and Knowledge Management}{October
17--21, 2022}{Atlanta, GA, USA}
\acmBooktitle{Proceedings of the 31st ACM International Conference on Information
and Knowledge Management (CIKM '22), October 17--21, 2022, Atlanta, GA, USA}
\acmPrice{15.00}
\acmDOI{10.1145/3511808.3557683}
\acmISBN{978-1-4503-9236-5/22/10}

\usepackage{multirow}
\usepackage{arydshln}
\usepackage{bm}
\usepackage{titlesec}

\begin{document}

\title[Rethinking Large-scale Pre-ranking System: Entire-chain Cross-domain Models]{Rethinking Large-scale Pre-ranking System: \\ Entire-chain Cross-domain Models}



\author{Jinbo Song}
\author{Ruoran Huang}
\author{Xinyang Wang}
\author{Wei Huang}
\email{songjinbo@jd.com}
\email{huangruoran1@jd.com}
\email{wangxinyang14@jd.com}
\email{huangwei19@jd.com}
\affiliation{%
  \institution{Marketing \& Commercialization Center, JD.com}
  \state{Beijing}
  \country{China}
}




\author{Qian Yu}
\author{Mingming Chen}
\author{Yafei Yao}
\author{Chaosheng Fan}
\email{yuqian81@jd.com}
\email{chenmingming1@jd.com}
\email{yaoyafei1@jd.com}
\email{fanchaosheng1@jd.com}
\affiliation{%
  \institution{Marketing \& Commercialization Center, JD.com}
  \state{Beijing}
  \country{China}
}




\author{Changping Peng}
\author{Zhangang Lin}
\author{Jinghe Hu}
\author{Jingping Shao}
\email{pengchangping@jd.com}
\email{linzhangang@jd.com}
\email{hujinghe@jd.com}
\email{shaojingping@jd.com}
\affiliation{%
  \institution{Marketing \& Commercialization Center, JD.com}
  \state{Beijing}
  \country{China}
}




\settopmatter{printacmref=false} 
\renewcommand\footnotetextcopyrightpermission[1]{} 
\pagestyle{plain} 

\begin{abstract}
Industrial systems such as recommender systems and online advertising, have been widely equipped with multi-stage architectures, which are divided into several cascaded modules, including matching, pre-ranking, ranking and re-ranking. As a critical bridge between matching and ranking, existing pre-ranking approaches mainly endure sample selection bias (SSB) problem owing to ignoring the entire-chain data dependence, resulting in sub-optimal performances. In this paper, we rethink pre-ranking system from the perspective of the entire sample space, and propose Entire-chain Cross-domain Models (ECM), which leverage samples from the whole cascaded stages to effectively alleviate SSB problem. Besides, we design a fine-grained neural structure named ECMM to further improve the pre-ranking accuracy. Specifically, we propose a cross-domain multi-tower neural network to comprehensively predict for each stage result, and introduce the sub-networking routing strategy with $L0$ regularization to reduce computational costs. Evaluations on real-world large-scale traffic logs demonstrate that our pre-ranking models outperform SOTA methods while time consumption is maintained within an acceptable level, which achieves better trade-off between efficiency and effectiveness. 
\end{abstract}


\maketitle

\section{Introduction}
\begin{figure*}
\setlength{\abovecaptionskip}{0.cm}
\setlength{\belowcaptionskip}{-0.cm}
\centering
\includegraphics[width=1.\linewidth]{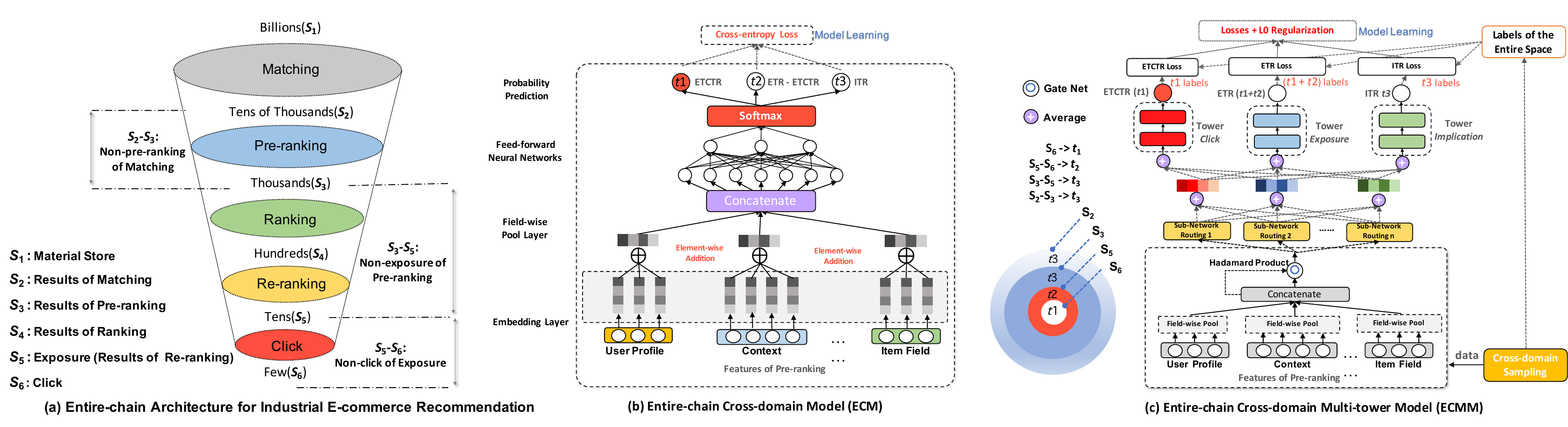}
\caption{(a) Illustration of cascade architecture for industrial  recommender system. (b) ECM is a coarse-grained pre-ranking multi-classfication model based on MLP, which trains cross-domain samples from the entire-chain data distribution. (c) ECMM is a fine-grained multi-tower structure by introducing the sub-network routing and $L0$ Regularization.}
\label{structures}
\end{figure*}

Online advertising system plays important roles in information service industry, and it contributes a lot to user experience and corporation income \cite{BCE}. To achieve a balance between effectiveness and efficiency, the multi-stage ranking architecture shown in Fig. \ref{structures}(a), including matching \cite{MOBIUS,Embedding}, pre-ranking \cite{COLD,FSCD}, ranking \cite{DIN,DIEN} and re-ranking \cite{Listwise,Re-Ranking}, is widely adopted. There exist abundant research works for stages of matching and ranking. Matching stage is to select thousands of relevant items from billion candidates, while ranking stage is responsible for finding out the most suitable candidates for users. Pre-ranking models, which are less complicated and originally derived from the ranking model, are normally adopted with representation-based architecture to ensure efficiency. Liu et al. \cite{Cascade} first propose cascade filtering framework in commercial search system, i.e., pre-ranking, ranking and re-ranking. Each candidate item would enter the next stage with the estimated probability from previous stage. Due to the limited online resource, it is challenging to balance the computational consumption and ranking quality which are both critical for user experience.


In industrial pre-ranking stage, vector-product based model is commonly employed as a rough information filtering to alleviate the computational burden of follow-up models in the ranking stage. Wang et.al \cite{COLD} propose a computational power cost-aware online deep pre-ranking model named COLD, whose main idea is to introduce squeeze-and-excitation blocks to calculate feature importance. Xu et al. \cite{FSCD} propose a feature selection method based on feature complexity and variational dropout named FSCD, to achieve a better trade-off between effectiveness and efficiency. FSCD training procedure is divided into two phases, namely pre-train and fine-tune. In pre-train phase, features are selected by jointly modeling feature complexity and performance. In fine-tune phase, only selected features from pre-train are used for subsequent training. In fact, as a a transitional stage between the matching and the ranking, models in pre-ranking directly determine the input quality of the downstream models
(following ranking and re-ranking models). However, aforementioned works focus on data distribution which comes from data distribution of exposure domain $S_{5}$ in Fig. \ref{structures}(a), ignoring the entire-chain data dependence from pre-ranking domain $S_{2}$ to clicking domain $S_{6}$. It inevitably causes Sample Selection Bias (SSB) problem \cite{Debiasing,ESMM}, which means that the model is trained on $S_{5}$ but makes inferences on $S_{2}$, resulting in sub-optimal performances. In this paper, we rethink pre-ranking system from a perspective of the entire-chain data space, and propose ECM as well as ECMM to effectively alleviate SSB problem. Experiments on  large-scale traffic logs demonstrate that our proposed pre-ranking models have achieved better efficiency and effectiveness.


\section{Methodology}
\textbf{2.1 Preliminaries}
\\
In industrial production environments, the entire-chain data distribution is defined as multiple domains with overlapping structures: $S_{1}\supset S_{2}\supset S_{3}\supset S_{4}\supset S_{5}\supset S_{6}$, whose definitions are shown in Fig. \ref{structures}(a). For e-commerce consumers, the most intuitive feeling comes from exposure domain $S_{5}$ and click domain $S_{6}$. We assume \boldsymbol{$x$} are sample features including user, item and cross information, $y5$ and $y6$ are binary labels with $y5=1$ or $y6=1$ indicating whether exposure or click event occurs, i.e. whether \boldsymbol{$x$} belong to $S_{5}$ or $S_{6}$, respectively. $y5\!=\!1\to y6\!=\!1$ reveals the sequential dependence of exposure and click labels, that there must be a preceding of exposure event when click event occurs. For an online pre-ranking system, the most concerned indicators are posterior Exposure-Through Rate (ETR) given that matching occurs and posterior Click-Through Rate (CTR) given that exposure occurs, which are estimated with $pETR=p(y5=1|\boldsymbol{x})$ and $pCTR=p(y6=1|y5=1, \boldsymbol{x})$, respectively. Since click event always occurs after exposure event, the actual click rate is calculated on post Exposure-Through\&Click-Through Rate (ETCTR) with $pETCTR = p(y6 = 1, y5 = 1|\boldsymbol{x})$ when considering the domains above the exposure domain $S_{5}$.

\noindent \textbf{2.2 SSB Problem in Pre-ranking System}
\label{SSB Problem in Pre-ranking System}

\noindent Given an observed dataset $\mathcal{D}_{obs}=\left \{ (\boldsymbol{x}_{i},y_{i}) \right \}_{i=1}^{N}$\}, and $\boldsymbol{x}_{i}$ is a feature vector of the sample
$i$ while $y_{i}$ is the corresponding label. Suppose $\hat{y}_{i}\in[0,1]$ indicates the prediction value and $y_{i}\in\{0,1\}$ indicates the ground truth of user $u$ on item $i$. Normally, the model is trained based on observed data by optimizing loss $\mathcal{L}(y_{i},\hat{y}_{i})$:
\begin{small}
\begin{equation}
\setlength\abovedisplayskip{-0.002cm}
\setlength\belowdisplayskip{-0.002cm}
\mathcal{L}(y_{i},\hat{y}_{i})=-y_{i}\text{log}\hat{y}_{i}-(1-y_{i})\text{log}(1-\hat{y}_{i})
\label{binary_loss}
\end{equation}
\end{small}

For current pre-ranking models, the objective function of training set can be formulated as:
\begin{small}
\begin{equation}
\setlength\abovedisplayskip{-0.05cm}
\setlength\belowdisplayskip{-0.05cm}
\begin{aligned}
\mathcal{L}_{\text{train}}(\mathbf{y} ^{obs},\hat{\mathbf{y}})=\frac{1}{\left | \mathcal{D}_{obs} \right | }
\sum_{\mathcal{D}_{obs} }\mathcal{L}(y_{i},\hat{y}_{i})
\end{aligned}
\end{equation}
\end{small}

\noindent where observed data $\mathcal{D}_{obs}$ is exposure domain.

However, in the real industrial scenario, pre-ranking models need to be evaluated over matching domain rather than exposure domain, where the former contains observed data $\mathcal{D}_{obs}$ and unobserved data $\mathcal{D}_{uno}$, and the latter only consists of observed data $\mathcal{D}_{obs}$. Then, the objective loss function $\mathcal{L}_{\text{test}}$ should be defined as:
\begin{small}
\begin{equation}
\setlength\abovedisplayskip{-0.05cm}
\setlength\belowdisplayskip{-0.0015cm}
\mathcal{L}_{\text{test}}(\mathbf{y}^{obs\cup uno},\hat{\mathbf{y}})
=\frac{1}{\left | \mathcal{D}_{obs} \cup \mathcal{D}_{uno} \right | }
\sum_{ \mathcal{D} }\mathcal{L}(y_{i},\hat{y}_{i})
\end{equation}
\end{small}
\normalfont
where $\mathcal{D}=\mathcal{D}_{uno}\cup \mathcal{D}_{obs}$. Generally speaking, estimator $\mathcal{M}$ is unbiased when the expectation of the estimated prediction inaccuracy over observed data $\mathcal{D}_{obs}$ equals to the prediction inaccuracy $\mathcal{L}_{test}$ over entire-chain data, i.e., $\text{Bias}^{\mathcal{M}}= |\mathbb{E}_{\mathcal{D}}[\mathcal{L}^{M}]-\mathcal{L}_{test}|= 0$, otherwise it is biased. 

However, $\mathbb{E}_{\mathcal{D}}[\mathcal{L}^{M}]$ is calculated on $\mathcal{D}_{obs}$ rather than $\mathcal{D}=\mathcal{D}_{uno}\cup \mathcal{D}_{obs}$, leading to  $\text{Bias}^{\mathcal{M}} \gg 0$. So if we only focus on exposure domain, i.e., $S_{5}$ in Fig. \ref{structures}(a), which contains exposure and click information, while ignoring unexposed samples above $S_{5}$, the model inevitably suffers from sample selection bias (SSB) due to the inconsistency between the distribution of test and training.

To alleviate SSB problem, one solution is treating unexposed samples as negative samples, i.e., Missing Not At Random (MNAR), but it makes unexposed samples underestimated and the model is unstable \cite{MNAR}. Another way is to adopt joint probability in ESMM to pre-ranking systems, we notice that ESMM actually alleviate the SSB between CTR ($S_{6}$) and ETR ($S_{5}$) to a certain extent. However, ESMM fails to account for the causes of the missing distribution above $S_{5}$, and lacks of modeling processes of non-pre-ranking of matching ($S_{2}-S_{3}$) and non-exposure of pre-ranking ($S_{3}-S_{5}$), which still results in $\text{Bias}^{\mathcal{M}} \gg 0$. 

\noindent \textbf{2.3 Proposed Approach}


\noindent \textbf{Learning Entire-chain Cross-domain Distribution.} In order to learn the entire-chain data distribution of the cascaded structures, distinct from the definition of upper-level in ESMM, pETR is redefined by using cross-domain information: $pETR=p(y5=1|\boldsymbol{x})=S_{5}/S_{2}$, which covers missing domains from $S_{5}$ (Exposure) to $S_{2}$ (Results of Matching). Then, two associated probabilities can be calculated as:
\begin{small}
\begin{equation}
\setlength\abovedisplayskip{0.05cm}
\setlength\belowdisplayskip{0.05cm}
\begin{aligned}
  &\underbrace{p(y6=1,y5=1|\boldsymbol{x})}_{pETCTR} =  \underbrace{p(y5=1|\boldsymbol{x})}_{pETR}\times   \underbrace{p(y6=1|y5=1,\boldsymbol{x})}_{pCTR}\\
&=\frac{Exposure(S_{5})}{Matching(S_{2})}\times \frac{Click(S_{6})}{Exposure(S_{5})}=\frac{Click(S_{6})}{Matching(S_{2})}
\label{Click_Matching}
\end{aligned}
\end{equation}
\end{small}
Eq. \ref{Click_Matching} tells us that by calculating associated probabilities of $pETR$ and $pCTR$, we can obtain 
the entirely cascaded data distribution, which addresses the SSB problem by sampling cross-domain data. 

Generally, labels of the exposure sample $S_{5}$ and the click sample $S_{6}$ are consistent with CTR ranking task. Considering that unexposed samples (above $S_{5}$) have no explicit user feedback, we can regard the whole cascaded process including stages of matching, pre-ranking, ranking and re-ranking as a unified behavior of user $u$. Then the click set $S_6$ is viewed as strong interest of $u$, and non-click of exposure set $S_{5}-S_{6}$ is viewed as weak interest while non-exposure of pre-ranking $S_{3}-S_{5}$ indicates that $u$ has no interest in it. Conceptually, the set of non-pre-ranking of matching $S_{2}-S_{3}$ also reflects an underlying interest preference. However, the objective of pre-ranking systems focus on user clicks and over-fined classification would damage the consistency of subsequent CTR task in ranking. Thereby, we merge $S_{3}-S_{5}$ and $S_{2}-S_{3}$ together as a unified learning goal named implication set domain. In this way, we transform SSB problem into a multi-classification task and obtain an Entire-chain Cross-domain (EC) paradigm of pre-ranking.


\noindent \textbf{ECM: Entire-chain Cross-domain Model.} Let $t1$, $t2$, $t3$ be learning objectives of click domain $S_{6}$, non-click of exposure domain $S_{5}-S_{6}$ and non-exposure of matching domain $(S_{3}-S_{5})\cup (S_{2}-S_{3})$, respectively. Then our targets are supposed to be rewritten as ETCTR, ETR-ETCTR and ITR (Implication-Through Rate), which are listed in Table \ref{Statistics of Datasets}. Naturally, we adopt the philosophy of multi-classification learning \cite{MMOE} and propose a multi-classification prediction strategy. As shown in Fig. \ref{structures}(b), we first propose an Entire-chain Cross-domain Model (ECM) to help the pre-ranking system to model the samples through feeding cross-domain data from the entire cascaded input space. ECM is based on basic neural network without complicated steel structures, and leverages \emph{softmax} operation to predict cross-domain labels. Let $T$ denote the number of multi-classification and the loss is the cross-entropy.



\noindent \textbf{ECMM: Entire-chain Cross-domain Multi-tower Model.} Although ECM is able to learn cross-domain information from different cascaded stages, it is still a coarse-grained network structure and cannot accurately model the data distribution of each domain. Meanwhile, multi-task learning is also perceived more cost-effective in training phase by applying parallel training mechanism \cite{Debiasing} than single \emph{softmax} function. Then we propose an Entire-chain Cross-domain Multi-tower Model (ECMM), which is illustrated in Fig. \ref{structures}(c). Compared with ECM, ECMM is a completely multi-task model that splits the whole neural network model into some forms of sub-networks, allowing different tasks to utilize different sub-networks. For the bottom-layer feature process, we integrate the gate network to preliminarily filter out ineffective feature vectors by controlling the gate switch: $E{}'=E\odot \sigma(W_{g}^{\top }E)$,

where $E$ donates the concatenated vector from bottom look-up table operations, $W_{g}$ is a mapping matrix and $\sigma$ means $\sigma(x)=1/(1+\text{exp}(-x))$. Meanwhile, to better alleviate the task conflicts and maintain the computation efficiency at serving time, we introduce sub-network routing by controlling the connection routing among the sub-networks of different layers. To reduce the number of parameters in our pre-ranking system, we adopt a weighted average with scalar coding variables as our final routing weight strategy \cite{SNR}. Subsequently, we construct three task-based structures according to different prediction objectives, on which towers of \emph{Click}, \emph{Exposure} and \emph{Implication} are built.

Different from \emph{softmax} cross-entropy in ECM, the loss of $\mathcal{L}_{\text{ECMM}}$ consists of $\mathcal{L}_{\text{ETCTR}}=\mathcal{L}(y6,t1)$, $\mathcal{L}_{\text{ETR}}=\mathcal{L}(y5,t1+t2)$ and $\mathcal{L}_{\text{ITR}}=\mathcal{L}(1-y5,t3)$, where each domain loss is calculated with an independent binary cross-entropy in Eq. \ref{binary_loss}. Considering limited computational resources, we also apply $L0$ regularization \cite{L0_Regularization} to reduce computational costs. Then the final objective function is:
\begin{small}
\begin{equation}
\setlength\abovedisplayskip{0.01cm}
\setlength\belowdisplayskip{-0.01cm}
\mathcal{L}_{\text{ECMM}}=\mathcal{L}_{\text{ETCTR}}+\mathcal{L}_{\text{ETR}}+\mathcal{L}_{\text{ITR}}+L0\; Regularization
\end{equation}
\end{small}

The multi-tower structure and sub-networks would inevitably increase the training and inference time. Here, we apply $L0$ regularization \cite{L0_Regularization} to reduce the model computational complexity by smoothing the distribution parameters of ECMM. Let $s$ be a continuous random variable and object variable $\boldsymbol{z}$ is learned by hard concrete distribution \cite{SNR} from Bernoulli distribution:
\begin{small}
\begin{equation}
\setlength\abovedisplayskip{0.01cm}
\setlength\belowdisplayskip{-0.01cm}
\begin{aligned}
s=\text{sigmoid}&((\text{log}(m)-\text{log}(1-m)+\text{log}(\alpha))/\beta), m\sim U(0,1)\\
\bar{s}&=s(\zeta-\gamma)+\gamma,\,\,\, z=\text{min}(1,\text{max}(\bar{s},0))
\label{Z_distribution}
\end{aligned}
\end{equation}
\end{small}

\noindent where $m$ is an uniform random variable, $\text{log}(\alpha)$ is a learnable distribution parameter, and $\beta$, $\gamma$, $\zeta$ are all hyper-parameters.



\section{Experiment}
\subsection{Experimental Setup}

\begin{table}
\centering\scriptsize
\setlength{\abovecaptionskip}{0.cm}
\setlength{\belowcaptionskip}{-0.cm}
\caption{Statistics of offline Datasets\protect\footnotemark[1], where \emph{M},\emph{B} means million and billion, respectively. } 

\setlength{\tabcolsep}{1.25mm}{
\begin{tabular}{ccccccc}
\hline
symbol & target &data description &stage& label & sample rate & size\\
\hline
$t1$ & ETCTR & click &$\bm{S}_{6}$& [1,0,0] & 100\% & 25.0 \emph{M}\\
$t2$ & ETR-ETCTR & non-click of exposure&$\bm{S}_{5}$-$\bm{S}_{6}$& [0,1,0] & 40\% & 1.10 \emph{B}\\
\multirow{2}{*}{$t3$}& \multirow{2}{*}{ITR} & non-exposure of pre-ranking &$\bm{S}_{3}$-$\bm{S}_{5}$& \multirow{2}{*}{[0,0,1]} & 1\% & 31.8 \emph{B}\\
 && non-pre-ranking of matching &$\bm{S}_{2}$-$\bm{S}_{3}$&  & 0.1\% & 162 \emph{B}\\
\hline 
\end{tabular}}
\label{Statistics of Datasets}
\vspace{-0.70cm}
\end{table}

\footnotetext[1]{https://github.com/songjinbo/ECMM.}



\textbf{Dataset.} Evaluations are mainly executed in JD online advertising system, covering 193 billion real-world e-commerce samples for 10 days. Because of a tremendous amount of samples in $S_{3}-S_{5}$ and $S_{2}-S_{3}$, we conduct the empirical random sampling to alleviate the computational burden and reduce excessive data imbalance. Dataset is listed in Table \ref{Statistics of Datasets}. \textbf{Baselines.} Our approaches are compared with the pre-ranking baselines of vector-product based model \cite{DSSM}, COLD \cite{COLD} and FSCD \cite{FSCD}. Although ESMM \cite{ESMM} is employed in the ranking stage and responsible for the Conversion Rate (CVR), we also compared with it due to some common structural properties. \textbf{Metrics.} AUC and GAUC \cite{DIN} are used as metrics to measure the ranking ability of models. Besides, we also measure the top-K Recall, to evaluate the recall ability of items that users are interested in. Ranking Consistency Score (RCS) \cite{RCS} is a ranking consistency metric for evaluating SSB which measures how well the pre-ranking model can select valuable items for the ranking stage. \textbf{Parameters.} For fairness, bottom-shared embeddings dimension is set to 8. All the models are trained using Adam with the learning rate as a tunable hyperparameter. For Vector-Product, COLD, FSCD and ESMM, we adopt the same structures as their works. The number of multi-classification $T$ is set to 3 according to Table \ref{Statistics of Datasets}. And for ECM, the hidden layers are adopted with sizes of 128, 64 and 32. For ECMM, we adopt two sub-layers of sub-routing networks with the sizes of 8 and 4. $\lambda=1e^{-5}$ is configured and the hyper-parameters for hard concrete distribution are grid search from the following range: $\beta \sim \left [ 0.5, 0.9 \right ] $, $\gamma\sim \left [ -1,-0.1 \right ]$, $\zeta \sim \left [ 1.1,2 \right ] $.
\subsection{Model Analysis}

\begin{table}
\centering\tiny
\setlength{\abovecaptionskip}{0.cm}
\setlength{\belowcaptionskip}{-0.cm}
\caption{\small{Offline evaluation results w.r.t. AUC, GAUC and Recall. }}
\setlength{\tabcolsep}{0.65mm}{
\begin{tabular}{cccccccccccc}
\hline
\multirow{3}{*}{Model}&
\multirow{3}{*}{AUC}&
\multirow{3}{*}{GAUC}&
\multicolumn{9}{c}{Recall@K} \\
\cline{5-12}
& & & \multicolumn{3}{c}{Recall@1} & \multicolumn{3}{c}{Recall@10} & \multicolumn{3}{c}{Recall@50} \\
& & & & exposure & click && exposure & click && exposure & click \\
\hline
COLD${}^{\ast}$ & 0.6268 & 0.5803 && 4.68\% & 7.30\% && 28.54\% & 38.90\% && 65.68\% & 75.36\%\\
FSCD${}^{\ast}$ & 0.6184 & 0.5708 && 2.43\% & 3.83\% && 21.24\% & 29.39\% && 60.59\% & 70.55\%\\
\hline

Vector-Product& 0.5945 & 0.5452 && 1.99\% & 2.57\% && 14.99\% & 17.60\% && 52.17\% & 55.72\%\\
COLD & 0.6107 & 0.5768 && 6.68\% & 9.69\% && 40.63\% & 50.17\% && 81.00\% & 85.81\%\\
FSCD & 0.6045 & 0.5712 && 5.88\% & 8.66\% && 38.01\% & 45.94\% && 79.01\% & 83.22\%\\
\hdashline
COLD(t1)${}^{\Im}$& 0.6076 & 0.5758 && 7.64\% & 10.87\% && 47.15\% & 54.28\% && 88.31\% & 90.70\%\\
FSCD(t1)${}^{\Im}$& 0.6022 & 0.5684 && 5.81\% & 7.99\% && 41.40\% & 48.03\% && 87.59\% & 88.89\%\\
ESMM(t1) & 0.6121 & 0.5776 && 8.27\% & 11.72\% && 49.29\% & 56.27\% && 89.30\% & 91.53\%\\
ECM(t1) & 0.6118 & 0.5790 && 8.12\% & 11.72\% && 48.81\% & 56.27\% && 89.19\% & 91.44\%\\
ECMM(t1) & 0.6143 & 0.5797 && 8.65\% & 12.12\% &&50.12\% & 56.41\% &&89.31\% &91.55\%\\
\hdashline
COLD(t2)${}^{\Im}$ & 0.5183 & 0.5432 && 8.72\% & 8.70\% && 51.45\% & 53.74\% && 88.84\% & 90.56\%\\
FSCD(t2)${}^{\Im}$ & 0.5128 & 0.5399 && 8.09\% & 7.84\% && 48.97\% & 50.53\% && 87.50\% & 89.12\%\\
ESMM(t2)& 0.5158 & 0.5452 && 8.96\% & 8.90\% && 52.42\% & 55.35\% && 89.47\% & 91.14\%\\
ECM(t2)& 0.5148 & 0.5450 && 8.93\% & 8.91\% && 52.27\% & 55.08\% && 89.43\% & 91.13\%\\
ECMM(t2)& 0.5369 & 0.5572 && 9.98\% & 11.72\% && 53.16\% & 56.79\% && 89.86\% & 91.31\%\\
\hline
\end{tabular}}
\flushleft
\footnotesize{Note: All models tests in exposure domain $S_{5}$. $\ast$ means that models are only trained on exposure data $S_{5}$, and $\Im$ represents that applying \emph{softmax} strategy in prediction.}


\label{offline_performance}
\vspace{-0.42cm}
\end{table}

\begin{figure}
\setlength{\abovecaptionskip}{0.cm}
\setlength{\belowcaptionskip}{-0.cm}
\centering
\includegraphics[width=0.8\linewidth]{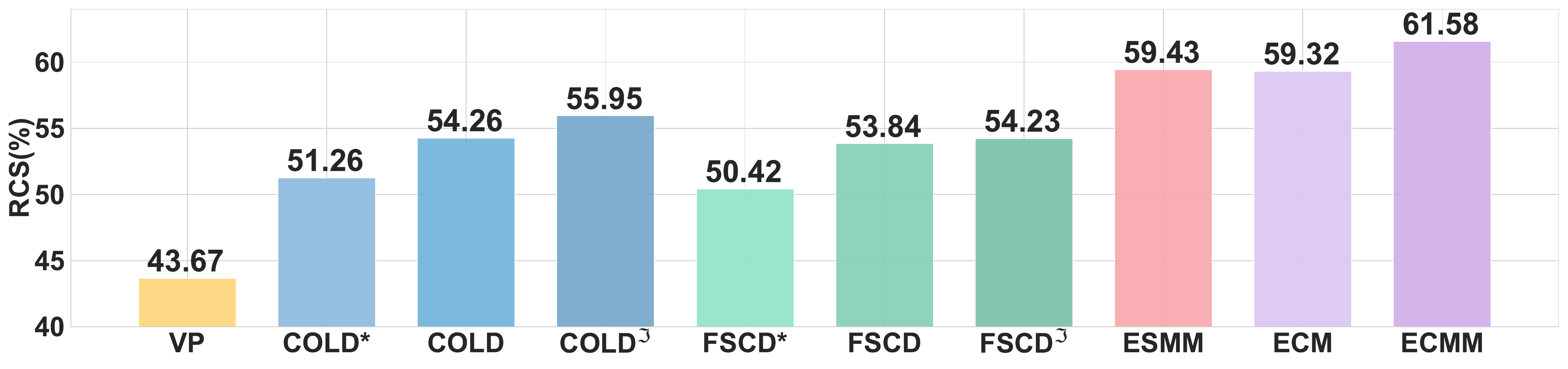}
\caption{Evaluation results w.r.t. RCS for all models. }
\label{RCS}
\vspace{-0.3cm}
\end{figure}
\noindent \textbf{Offline Performance Comparison.} Considering primary goals of the pre-ranking system are predicting user click actions and items in the exposure domain, we only report $t1$ and $t2$ results.  1) To validate whether our proposed strategies can alleviate SSB problem, we train SOTA methods of vanilla COLD and FSCD with exposure data ($S_{5}$, marked as ${}^{\ast}$) and entire-chain data ($S_{2}$ to $S_{6}$). From Table \ref{offline_performance}, we see that although COLD${}^{\ast}$ and FSCD${}^{\ast}$ w.r.t. AUC/GAUC increase slightly compared with their entire-chain training models, the metrics of Recall@K suffer a large loss, seriously dropping the subsequent online ranking performances. It is worth noting that one percent increasing of Recall can bring great benefits to advertising system. This phenomenon can be summarized as sampling bias due to missing entire-chain data in Section \ref{SSB Problem in Pre-ranking System}. 2) Meanwhile, to further solve SSB issue, we apply EC paradigm and \emph{softmax} strategy on COLD and FSCD models to conduct the heuristic experiments. Although the performance of COLD (FSCD) on AUC decreases slightly, the metric Recall has achieved significant improvements on exposure and click of approximately 14.79\% (6.19\%) and 10.18\% (3.52\%) compared with COLD (FSCD). 3) ECMM consistently yields the best performance and outperforms other baselines w.r.t. AUC/GAUC and Recall in Table \ref{offline_performance}. As a brief version of ECMM, ECM maintains a comparable AUC/GAUC with ECMM. 4) Even compared with SOTA methods of COLD and FSCD, ECM still reveals its competitiveness on $t1$, indicating that the entire-chain cross-domain paradigm indeed advances the performance. 5) ESMM gets a comparable result with ECM, and this attributes to that both of them have similar structures except for output forms. 

To verify the ranking consistency, we also report RCS \cite{RCS} metric in Fig. \ref{RCS}, where the higher score means better performance. 1) Note that COLD${}^{*}$<COLD and FSCD${}^{*}$<FSCD, from which we draw conclusions that our proposed EC paradigm effectively alleviate SSB problem. 2) From COLD<COLD${}^{\Im}$ and FSCD<FSCD${}^{\Im}$, we see that ECM-based models can further boost the accuracy by eliminating the gaps of sample selection bias with multi-classification prediction strategy. 3) ECMM which applies the fine-grained multi-tower and sub-networking structures outperforms other SOTA methods.


\begin{table}
\centering\tiny
\setlength{\abovecaptionskip}{0.cm}
\setlength{\belowcaptionskip}{-0.cm}
\caption{\small{Online test of pre-ranking systems w.r.t. different models.}}
\setlength{\tabcolsep}{0.25cm}{\begin{tabular}{cccccc}
\hline
Model & QPS(full CPU) & RT(TP99) & CPU & CTR & RPM\\
\hline
Vector-Product & 12243 & 0.874ms & 24\% & $-$ & $-$\\
COLD & 2122 & 1.823ms & 58\% & 2.3\% & 2.5\%\\
FSCD & 3763 & 1.423ms & 47\% & 2.1\% & 2.4\%\\
ESMM & 1608 & 2.301ms & 74\% & 3.8\% & 5.1\%\\
ECM${}^{*}$ & 1861 & 2.194ms & 66\% & 4.1\% & 5.0\%\\
ECMM${}^{*}$ & 1613 & 2.297ms & 73\% & 4.7\% & 5.6\%\\
\hline
\label{online_reults}
\end{tabular}}
\vspace{-0.2cm}
\end{table}

\begin{table}
\vspace{-0.5cm}
\centering\tiny
\setlength{\abovecaptionskip}{0.cm}
\setlength{\belowcaptionskip}{-0.cm}
\caption{\small{Ablation study of proposed pre-ranking models.}}
\setlength{\tabcolsep}{0.45cm}{\begin{tabular}{cccc}
\hline
Model& QPS & RT(TP99) & GAUC\\
\hline
ECM(half features)& 1977 & 2.018ms & 0.5740\\
ECM(all features) & 1861 & 2.194ms & 0.5790\\
ECMM(half features) & 1880 & 2.116ms & 0.5758\\
ECMM(all features) & 1613 & 2.297ms & 0.5797\\
ECMM/wo $L0$ & 1226 & 2.403ms & 0.5806\\
ECMM/w $t4$ & 914 & 2.615ms & 0.5895\\

\hline
\end{tabular}}
\flushleft
\footnotesize{Note: Symbol wo $L0$ means the model without adopting $L0$ regularization, and symbol w $t4$ means that using $S_{2}-S_{3}$ as the fourth tower.}
\label{Ablation}
\vspace{-0.5cm}
\end{table}


\noindent \textbf{Industrial Online Results.} In the online A/B test system, each model is trained over 200 billion real-world e-commerce samples. For the online evaluation, metrics including QPS (Queries Per Seconds), RT (Return Time), RPM (Revenue Per Mille) and CPU consumption are adopted. Table \ref{online_reults} shows 30 continuous normal days of mobile real-world requests, from which we see that proposed ECM achieves 4.1\% CTR and 5.1\% RPM improvement, while improvements turns to be 4.7\% and 5.6\% RPM of ECMM. Benefited 
from simple structures, RT and QPS of Vector-Product are far superior to other models. Meanwhile, we find that compared with COLD and FSCD, ECMM has gained approximate double performance improvements by costing acceptable time.

\noindent \textbf{Ablation Study.} We conduct ablation experiments on ECM and ECMM, then report user click actions ($t1$) results. Experiments conducted in offline datasets are shown in Table \ref{Ablation}, from which the following observations are made: 1) Although only a half of the input features which are selected according to \cite{COLD} are preserved, the model can still maintain a good predicting ability, indicating that our proposed approach effectively captures the data distribution and boost pre-ranking performances. 2) When removing $L0$ regularization, the time consumption increases while the performance has no clear changes. It reveals that applying $L0$ regularization actually reduces the computational consumption. 3) Adding $t4$ tower cannot bring further desirable benefits, in contrast, the increasing of parameters causes more time overhead. Empirically speaking, the architecture of three towers are proper for the pre-ranking task.

\section{Conclusion}
In this paper, we rethink pre-ranking system from the perspective of entire cascaded structures, and propose Entire-chain Cross-domain Models (ECM) to alleviate SSB problem, which train samples from entire-chain domains. Moreover, we propose a cross-domain multi-tower neural network named ECMM based on multi-task learning to comprehensively predict for each stage result. Besides, we introduce the sub-networking routing strategy with $L0$ regularization to reduce computational costs. Extensive experiments in the real-world industrial datasets demonstrate that our pre-ranking models outperform SOTA methods while time consumption is maintained within an acceptable level. The proposed model has been serving the main traffic in our online advertising system. 

\bibliographystyle{ACM-Reference-Format}
\bibliography{ms}


\begin{thebibliography}{18}


\ifx \showCODEN    \undefined \def \showCODEN     #1{\unskip}     \fi
\ifx \showDOI      \undefined \def \showDOI       #1{#1}\fi
\ifx \showISBNx    \undefined \def \showISBNx     #1{\unskip}     \fi
\ifx \showISBNxiii \undefined \def \showISBNxiii  #1{\unskip}     \fi
\ifx \showISSN     \undefined \def \showISSN      #1{\unskip}     \fi
\ifx \showLCCN     \undefined \def \showLCCN      #1{\unskip}     \fi
\ifx \shownote     \undefined \def \shownote      #1{#1}          \fi
\ifx \showarticletitle \undefined \def \showarticletitle #1{#1}   \fi
\ifx \showURL      \undefined \def \showURL       {\relax}        \fi
\providecommand\bibfield[2]{#2}
\providecommand\bibinfo[2]{#2}
\providecommand\natexlab[1]{#1}
\providecommand\showeprint[2][]{arXiv:#2}

\bibitem[Ai et~al\mbox{.}(2018)]%
        {Listwise}
\bibfield{author}{\bibinfo{person}{Qingyao Ai}, \bibinfo{person}{Keping Bi}, \bibinfo{person}{Jiafeng Guo}, {and} \bibinfo{person}{W.~Bruce Croft}.} \bibinfo{year}{2018}\natexlab{}.
\newblock \showarticletitle{Learning a Deep Listwise Context Model for Ranking Refinement}.
\newblock \bibinfo{journal}{\emph{The 41st International ACM SIGIR Conference on Research Development in Information Retrieval}} (\bibinfo{year}{2018}), \bibinfo{pages}{135–144}.
\newblock
\showISBNx{9781450356572}


\bibitem[Fan et~al\mbox{.}(2019)]%
        {MOBIUS}
\bibfield{author}{\bibinfo{person}{Miao Fan}, \bibinfo{person}{Jiacheng Guo}, \bibinfo{person}{Shuai Zhu}, \bibinfo{person}{Shuo Miao}, \bibinfo{person}{Mingming Sun}, {and} \bibinfo{person}{Ping Li}.} \bibinfo{year}{2019}\natexlab{}.
\newblock \showarticletitle{MOBIUS: Towards the Next Generation of Query-Ad Matching in Baidu's Sponsored Search}.
\newblock \bibinfo{journal}{\emph{Proceedings of the 25th ACM SIGKDD International Conference on Knowledge Discovery \& Data Mining}} (\bibinfo{year}{2019}), \bibinfo{pages}{2509–2517}.
\newblock


\bibitem[Gu et~al\mbox{.}(2022)]%
        {RCS}
\bibfield{author}{\bibinfo{person}{Siyu Gu}, \bibinfo{person}{Xiang-Rong Sheng}, \bibinfo{person}{Biye Jiang}, \bibinfo{person}{Siyuan Lou}, \bibinfo{person}{Shuguang Han}, \bibinfo{person}{Hongbo Deng}, {and} \bibinfo{person}{Bo Zheng}.} \bibinfo{year}{2022}\natexlab{}.
\newblock \showarticletitle{On Ranking Consistency of Pre-ranking Stage}.
\newblock \bibinfo{journal}{\emph{arXiv preprint arXiv:2205.01289}} (\bibinfo{year}{2022}).
\newblock


\bibitem[Huang et~al\mbox{.}(2020)]%
        {Embedding}
\bibfield{author}{\bibinfo{person}{Jui-Ting Huang}, \bibinfo{person}{Ashish Sharma}, \bibinfo{person}{Shuying Sun}, \bibinfo{person}{Li Xia}, \bibinfo{person}{David Zhang}, \bibinfo{person}{Philip Pronin}, \bibinfo{person}{Janani Padmanabhan}, \bibinfo{person}{Giuseppe Ottaviano}, {and} \bibinfo{person}{Linjun Yang}.} \bibinfo{year}{2020}\natexlab{}.
\newblock \showarticletitle{Embedding-based retrieval in facebook search}.
\newblock \bibinfo{journal}{\emph{Proceedings of the 26th ACM SIGKDD International Conference on Knowledge Discovery Data Mining}} (\bibinfo{year}{2020}), \bibinfo{pages}{2553--2561}.
\newblock


\bibitem[Huang et~al\mbox{.}(2013)]%
        {DSSM}
\bibfield{author}{\bibinfo{person}{Po{-}Sen Huang}, \bibinfo{person}{Xiaodong He}, \bibinfo{person}{Jianfeng Gao}, \bibinfo{person}{Li Deng}, \bibinfo{person}{Alex Acero}, {and} \bibinfo{person}{Larry~P. Heck}.} \bibinfo{year}{2013}\natexlab{}.
\newblock \showarticletitle{Learning deep structured semantic models for web search using clickthrough data}.
\newblock \bibinfo{journal}{\emph{22nd {ACM} International Conference on Information and Knowledge Management (CIKM)}} (\bibinfo{year}{2013}), \bibinfo{pages}{2333--2338}.
\newblock


\bibitem[Liu et~al\mbox{.}(2017)]%
        {Cascade}
\bibfield{author}{\bibinfo{person}{Shichen Liu}, \bibinfo{person}{Fei Xiao}, \bibinfo{person}{Wenwu Ou}, {and} \bibinfo{person}{Luo Si}.} \bibinfo{year}{2017}\natexlab{}.
\newblock \showarticletitle{Cascade Ranking for Operational E-commerce Search}.
\newblock \bibinfo{journal}{\emph{Proceedings of the 23rd {ACM} {SIGKDD} International Conference on Knowledge Discovery and Data Mining}} (\bibinfo{year}{2017}), \bibinfo{pages}{1557--1565}.
\newblock


\bibitem[Louizos et~al\mbox{.}(2018)]%
        {L0_Regularization}
\bibfield{author}{\bibinfo{person}{Christos Louizos}, \bibinfo{person}{Max Welling}, {and} \bibinfo{person}{Diederik~P. Kingma}.} \bibinfo{year}{2018}\natexlab{}.
\newblock \showarticletitle{Learning Sparse Neural Networks through L\({}_{\mbox{0}}\) Regularization}.
\newblock \bibinfo{journal}{\emph{Proceedings of the 6th International Conference on Learning Representations (ICLR)}} (\bibinfo{year}{2018}).
\newblock


\bibitem[Ma et~al\mbox{.}(2019)]%
        {SNR}
\bibfield{author}{\bibinfo{person}{Jiaqi Ma}, \bibinfo{person}{Zhe Zhao}, \bibinfo{person}{Jilin Chen}, \bibinfo{person}{Ang Li}, \bibinfo{person}{Lichan Hong}, {and} \bibinfo{person}{Ed~H Chi}.} \bibinfo{year}{2019}\natexlab{}.
\newblock \showarticletitle{SNR: sub-network routing for flexible parameter sharing in multi-task learning}.
\newblock \bibinfo{journal}{\emph{Proceedings of the AAAI Conference on Artificial Intelligence}} \bibinfo{volume}{33}, \bibinfo{number}{01} (\bibinfo{year}{2019}), \bibinfo{pages}{216--223}.
\newblock


\bibitem[Ma et~al\mbox{.}(2021)]%
        {FSCD}
\bibfield{author}{\bibinfo{person}{Xu Ma}, \bibinfo{person}{Pengjie Wang}, \bibinfo{person}{Hui Zhao}, \bibinfo{person}{Shaoguo Liu}, \bibinfo{person}{Chuhan Zhao}, \bibinfo{person}{Wei Lin}, \bibinfo{person}{Kuang-Chih Lee}, \bibinfo{person}{Jian Xu}, {and} \bibinfo{person}{Bo Zheng}.} \bibinfo{year}{2021}\natexlab{}.
\newblock \showarticletitle{Towards a Better Tradeoff between Effectiveness and Efficiency in Pre-Ranking: A Learnable Feature Selection Based Approach}.
\newblock \bibinfo{journal}{\emph{Proceedings of the 44th International ACM SIGIR Conference on Research and Development in Information Retrieval}} (\bibinfo{year}{2021}), \bibinfo{pages}{2036–2040}.
\newblock
\showISBNx{9781450380379}


\bibitem[Ma et~al\mbox{.}(2018)]%
        {ESMM}
\bibfield{author}{\bibinfo{person}{Xiao Ma}, \bibinfo{person}{Liqin Zhao}, \bibinfo{person}{Guan Huang}, \bibinfo{person}{Zhi Wang}, \bibinfo{person}{Zelin Hu}, \bibinfo{person}{Xiaoqiang Zhu}, {and} \bibinfo{person}{Kun Gai}.} \bibinfo{year}{2018}\natexlab{}.
\newblock \showarticletitle{Entire Space Multi-Task Model: An Effective Approach for Estimating Post-Click Conversion Rate}.
\newblock  (\bibinfo{year}{2018}), \bibinfo{pages}{1137--1140}.
\newblock


\bibitem[Pei et~al\mbox{.}(2019)]%
        {Re-Ranking}
\bibfield{author}{\bibinfo{person}{Changhua Pei}, \bibinfo{person}{Yi Zhang}, \bibinfo{person}{Yongfeng Zhang}, \bibinfo{person}{Fei Sun}, \bibinfo{person}{Xiao Lin}, \bibinfo{person}{Hanxiao Sun}, \bibinfo{person}{Jian Wu}, \bibinfo{person}{Peng Jiang}, \bibinfo{person}{Junfeng Ge}, \bibinfo{person}{Wenwu Ou}, {and} \bibinfo{person}{Dan Pei}.} \bibinfo{year}{2019}\natexlab{}.
\newblock \showarticletitle{Personalized Re-Ranking for Recommendation}.
\newblock \bibinfo{journal}{\emph{Proceedings of the 13th ACM Conference on Recommender Systems}} (\bibinfo{year}{2019}), \bibinfo{pages}{3–11}.
\newblock
\showISBNx{9781450362436}


\bibitem[Schnabel et~al\mbox{.}(2016)]%
        {MNAR}
\bibfield{author}{\bibinfo{person}{Tobias Schnabel}, \bibinfo{person}{Adith Swaminathan}, \bibinfo{person}{Ashudeep Singh}, \bibinfo{person}{Navin Chandak}, {and} \bibinfo{person}{Thorsten Joachims}.} \bibinfo{year}{2016}\natexlab{}.
\newblock \showarticletitle{Recommendations as Treatments: Debiasing Learning and Evaluation}.
\newblock \bibinfo{journal}{\emph{Proceedings of the 33rd International Conference on International Conference on Machine Learning - Volume 48}} (\bibinfo{year}{2016}), \bibinfo{pages}{1670–1679}.
\newblock


\bibitem[Tang et~al\mbox{.}(2020)]%
        {MMOE}
\bibfield{author}{\bibinfo{person}{Hongyan Tang}, \bibinfo{person}{Junning Liu}, \bibinfo{person}{Ming Zhao}, {and} \bibinfo{person}{Xudong Gong}.} \bibinfo{year}{2020}\natexlab{}.
\newblock \showarticletitle{Progressive layered extraction (ple): A novel multi-task learning model for personalized recommendations}.
\newblock \bibinfo{journal}{\emph{Proceedings of the Fourteenth ACM Conference on Recommender Systems}} (\bibinfo{year}{2020}), \bibinfo{pages}{269--278}.
\newblock


\bibitem[Wang et~al\mbox{.}(2018)]%
        {BCE}
\bibfield{author}{\bibinfo{person}{Jizhe Wang}, \bibinfo{person}{Pipei Huang}, \bibinfo{person}{Huan Zhao}, \bibinfo{person}{Zhibo Zhang}, \bibinfo{person}{Binqiang Zhao}, {and} \bibinfo{person}{Dik~Lun Lee}.} \bibinfo{year}{2018}\natexlab{}.
\newblock \showarticletitle{Billion-scale Commodity Embedding for E-commerce Recommendation in Alibaba}.
\newblock \bibinfo{journal}{\emph{Proceedings of the 24th {ACM} {SIGKDD} International Conference on Knowledge Discovery {\&} Data Mining}} (\bibinfo{year}{2018}), \bibinfo{pages}{839--848}.
\newblock


\bibitem[Wang et~al\mbox{.}(2020)]%
        {COLD}
\bibfield{author}{\bibinfo{person}{Zhe Wang}, \bibinfo{person}{Liqin Zhao}, \bibinfo{person}{Biye Jiang}, \bibinfo{person}{Guorui Zhou}, \bibinfo{person}{Xiaoqiang Zhu}, {and} \bibinfo{person}{Kun Gai}.} \bibinfo{year}{2020}\natexlab{}.
\newblock \showarticletitle{{COLD:} Towards the Next Generation of Pre-Ranking System}.
\newblock \bibinfo{journal}{\emph{CoRR}}  \bibinfo{volume}{abs/2007.16122} (\bibinfo{year}{2020}).
\newblock


\bibitem[Zhang et~al\mbox{.}(2020)]%
        {Debiasing}
\bibfield{author}{\bibinfo{person}{Wenhao Zhang}, \bibinfo{person}{Wentian Bao}, \bibinfo{person}{Xiao{-}Yang Liu}, \bibinfo{person}{Keping Yang}, \bibinfo{person}{Quan Lin}, \bibinfo{person}{Hong Wen}, {and} \bibinfo{person}{Ramin Ramezani}.} \bibinfo{year}{2020}\natexlab{}.
\newblock \showarticletitle{Large-scale Causal Approaches to Debiasing Post-click Conversion Rate Estimation with Multi-task Learning}.
\newblock \bibinfo{journal}{\emph{Proceedings of The Web Conference (WWW)}} (\bibinfo{year}{2020}), \bibinfo{pages}{2775--2781}.
\newblock


\bibitem[Zhou et~al\mbox{.}(2019)]%
        {DIEN}
\bibfield{author}{\bibinfo{person}{Guorui Zhou}, \bibinfo{person}{Na Mou}, \bibinfo{person}{Ying Fan}, \bibinfo{person}{Qi Pi}, \bibinfo{person}{Weijie Bian}, \bibinfo{person}{Chang Zhou}, \bibinfo{person}{Xiaoqiang Zhu}, {and} \bibinfo{person}{Kun Gai}.} \bibinfo{year}{2019}\natexlab{}.
\newblock \showarticletitle{Deep Interest Evolution Network for Click-through Rate Prediction}.
\newblock \bibinfo{journal}{\emph{Proceedings of the Thirty-Third AAAI Conference on Artificial Intelligence and Thirty-First Innovative Applications of Artificial Intelligence Conference and Ninth AAAI Symposium on Educational Advances in Artificial Intelligence}}, Article \bibinfo{articleno}{729} (\bibinfo{year}{2019}), \bibinfo{numpages}{8}~pages.
\newblock
\showISBNx{978-1-57735-809-1}


\bibitem[Zhou et~al\mbox{.}(2018)]%
        {DIN}
\bibfield{author}{\bibinfo{person}{Guorui Zhou}, \bibinfo{person}{Xiaoqiang Zhu}, \bibinfo{person}{Chengru Song}, \bibinfo{person}{Ying Fan}, \bibinfo{person}{Han Zhu}, \bibinfo{person}{Xiao Ma}, \bibinfo{person}{Yanghui Yan}, \bibinfo{person}{Junqi Jin}, \bibinfo{person}{Han Li}, {and} \bibinfo{person}{Kun Gai}.} \bibinfo{year}{2018}\natexlab{}.
\newblock \showarticletitle{Deep Interest Network for Click-Through Rate Prediction}.
\newblock \bibinfo{journal}{\emph{Proceedings of the 24th {ACM} {SIGKDD} International Conference on Knowledge Discovery {\&} Data Mining, {KDD}}} (\bibinfo{year}{2018}), \bibinfo{pages}{1059--1068}.
\newblock


\end{thebibliography}


\end{document}